\documentclass[conference]{IEEEtran}
\IEEEoverridecommandlockouts
\usepackage{cite}
\usepackage{amsmath,amssymb,amsfonts}
\usepackage{algorithmic}
\usepackage{graphicx}
\usepackage{textcomp}
\usepackage{xcolor}
\usepackage{url}
\usepackage{listings}
\usepackage{booktabs} 
\usepackage{multirow} 
\usepackage{float}
\usepackage{hyperref}
\usepackage{todonotes}
\def\BibTeX{{\rm B\kern-.05em{\sc i\kern-.025em b}\kern-.08em
    T\kern-.1667em\lower.7ex\hbox{E}\kern-.125emX}}
\begin{document}

\title{LLM-Powered Intent-Based Categorization of
Phishing Emails \thanks{This research was funded by the European Union as part of the Horizon Europe project SYNAPSE (GA No. 101120853). Views and opinions expressed are, however, those of the author(s) only and do not necessarily reflect those of the European Union. Neither the European Union nor the granting authority can be held responsible for them.}\\

}

\author{\IEEEauthorblockN{Even Eilertsen}
\IEEEauthorblockA{\textit{University of Oslo} \\
Oslo, Norway \\
\texttt{eveneil@ifi.uio.no}}
\and
\IEEEauthorblockN{ Vasileios Mavroeidis}
\IEEEauthorblockA{\textit{University of Oslo} \\
Oslo, Norway \\
\texttt{vasileim@ifi.uio.no}}
\and
\IEEEauthorblockN{ Gudmund Grov}
\IEEEauthorblockA{\textit{Norwegian Defence Research Establishment (FFI)} \\ \& \textit{University of Oslo} \\
Kjeller, Norway \\
\texttt{Gudmund.Grov@ffi.no}}
}

\maketitle

\begin{abstract}
Phishing attacks remain a significant threat to modern cybersecurity, as they successfully deceive both humans and the defense mechanisms intended to protect them. Traditional detection systems primarily focus on email metadata that users cannot see in their inboxes. Additionally, these systems struggle with phishing emails, which experienced users can often identify empirically by the text alone. This paper investigates the practical potential of Large Language Models (LLMs) to detect these emails by focusing on their intent. In addition to the binary classification of phishing emails, the paper introduces an intent-type taxonomy, which is operationalized by the LLMs to classify emails into distinct categories and, therefore, generate actionable threat information. To facilitate our work, we have curated publicly available datasets into a custom dataset containing a mix of legitimate and phishing emails. Our results demonstrate that existing LLMs are capable of detecting and categorizing phishing emails, underscoring their potential in this domain.

\end{abstract}

\begin{IEEEkeywords}
cybersecurity, email security, phishing detection, large language models, AI, cyber threat information
\end{IEEEkeywords}

\section{Introduction}\label{sec:intro}
Phishing is a well-known attack technique dating back to at least the 1990s \cite{rekouche2011early}. As the use of the internet has continued to grow, so have the assets accessible online. In today's digital world, most businesses and organizations are connected to the internet, resulting in a substantial volume of email communication that malicious actors can exploit.

Phishing emails remain a prevalent threat \cite{alkhalil2021phishing}, as the majority of successful cyberattacks originate from phishing campaigns \cite{cofense2024emailsecurity,microsoft2024digitaldefense}. Many email defense mechanisms against phishing attacks focus on metadata, information around used protocols, and data besides the subject and body text fields within the email \cite{cofense2024emailsecurity}. Although such approaches have been successful in detecting phishing emails, other emails that experienced or trained users can easily identify as phishing by simply reading the text still evade detection. With this in mind, our hypothesis is as follows:
\begin{quote}\it
By addressing the language and intent of emails, LLMs can detect phishing in a manner that complements existing metadata-based detection techniques. 
\end{quote}
\emph{Large Language Models} (LLMs) have been shown to exhibit knowledge in this area, and this paper explores to what extent LLMs can act as "the experienced users" to detect phishing intent, both with inherent knowledge and through  \emph{in-context learning} using one or more examples.

Different types of phishing emails exist, each with a distinct intent, as characterized by various MITRE ATT\&CK techniques \cite{MITRE_ATTCK}. For example, the intent behind an untargeted mass phishing campaign typically differs significantly from that of a targeted spearphishing email, which contains personalized information about the victim. Exploring both in-context learning and phishing categories, this paper addresses the following five research questions:
\begin{description}
    \item[\textbf{(RQ1)}] To what extent can LLMs infer intent in emails and use that as a factor for phishing detection?
    \item[\textbf{(RQ2)}] To what extent is knowledge inherent in LLMs, and to what degree do examples in a few-shot learning setting help with detection?
    \item[\textbf{(RQ3)}] To what degree are LLMs able to explain and justify their reasoning?
    \item[\textbf{(RQ4)}] To what degree can LLMs differentiate between different types of phishing categories?
    \item[\textbf{(RQ5)}] To what degree does the contextual knowledge provided by the phishing categories help to identify phishing emails?
\end{description}

In addition to addressing the research questions, the contributions of the paper are as follows.
Based on the MITRE ATT\&CK framework \cite{MITRE_ATTCK}, we populate a taxonomy of phishing intent and use it to enrich a curated dataset\footnote {https://github.com/Rub3cula/Datasets/blob/main/100EmailsDataset.csv} of phishing emails. We then design a set of prompts and evaluate them under two settings. In the zero-shot approach, the prompt is presented with the email alone, without any examples of desired outputs. In the few-shot approach, the prompt includes example emails paired with correct labels to guide the model. This study evaluates multiple LLMs to assess their effectiveness in detecting phishing intent, revealing mixed results across models when using in-context learning.

\section{Related Work}\label{sec:background}

In the context of cybersecurity, defense against phishing attacks can be broadly categorised into two types: technical defenses and non-technical defenses \cite{brickley2021comparative}. Non-technical defenses primarily focus on educating potential targets—typically email recipients—through methods such as training courses and simulated phishing tests. These initiatives aim to build user awareness and resilience by teaching individuals how to recognize and respond to phishing attempts. In contrast, technical defenses play a critical role in securing email platforms through automated detection and prevention mechanisms.

In \cite{hazell2023large}, the author argues that LLMs can reduce the workload and skill barrier required to create high-quality, targeted phishing emails. The research indicates that certain detection mechanisms can be circumvented by carefully crafting phishing emails using LLMs through a method known as prompt engineering, which involves adjusting prompts to produce specific responses or outcomes. 
The paper proposes either restricting the functionality of advanced models or implementing traceability to prevent their misuse in malicious contexts. In addition, the author proposes an LLM-based defensive system in which the LLMs themselves can detect phishing emails, a crucial development given the strong indicators that LLMs will continue to improve, potentially enabling more sophisticated phishing attack campaigns. Phishing detection systems should consider the capability of prompt engineering to bypass content filters quickly \cite{hazell2023large}.

The authors in \cite{Heiding2024devising} provide empirical evidence that using LLMs to create phishing emails can achieve a greater incentivizing success rate than existing phishing emails gathered from online archives. Although LLMs have not outperformed manually written emails using a framework, phishing emails that are empowered by both LLMs and humans have achieved the best results. The authors developed phishing emails employing spearphishing techniques, integrating contextually relevant information tailored to specific targets. Although the primary objective of the paper was to examine the construction of such emails, it also proposed approaches for LLMs in phishing detection. In particular, the authors highlighted the importance of analyzing communicative intent as a potential differentiator between legitimate marketing content and malicious phishing attempts. Furthermore, \cite{hazell2023large} demonstrated how LLMs could create cost-effective and scalable spearphishing campaigns.

In addition to the subject line and body content, publicly available phishing datasets often include additional metadata, such as IP addresses and authentication protocol logs. Existing detection algorithms frequently leverage sender authentication mechanisms—such as SPF, DKIM, and DMARC \cite{derouet2016fighting}—which are commonly employed by traditional machine learning-based email security solutions. For example, datasets like SpamAssassin and the Anti-Phishing Working Group (APWG) provide IP addresses, domain information, and authentication results. Such data plays an essential role in research focused on analyzing the overall characteristics of emails. Furthermore, many phishing emails resemble poorly constructed spam messages, making them easier for users to identify and disregard.

Our paper primarily focuses on analyzing the intent of emails by examining only the subject line and body content, thereby simulating the way a typical user perceives an email. This approach is particularly valuable in scenarios where traditional detection mechanisms fail, allowing phishing messages to bypass security filters and reach users' inboxes.

ChatSpamDetector \cite{koide2024chatspamdetector} is a recent example in which LLMs have demonstrated strong performance in phishing detection, utilizing recent datasets and real-world emails to achieve an accuracy of 99.7\%. This significantly outperforms baseline systems and other traditional models. Despite these promising results, the approach is not intended to fully replace existing solutions. Deploying commercial LLMs—such as OpenAI's GPT-4o—at scale remains both cost-prohibitive and potentially non-compliant with privacy best practices \cite{yao2024survey}. 
Non-technical approaches focus on educating users on how to identify phishing emails \cite{jampen2020don}. Numerous studies utilize publicly available datasets to conduct their experiments. Many phishing emails included in these datasets might be considered unsophisticated attempts; however, due to varying email security configurations, they can still occasionally end up in users' inboxes  \cite{PhishingEmailDataset2024}. 
Recently, there has been significant progress in the field of LLMs, particularly in text reasoning tasks and zero-shot learning \cite{kojima2022large}. ChatSpamDetector used prompts to instruct LLMs on how to perform detection tasks effectively.

\section{An intent-type Phishing taxonomy}
The taxonomy used in this work is derived from the MITRE ATT\&CK Technique T1566 for phishing \cite{mitreT1566_002}, as presented in Table I. We adopt the sub-techniques defined by ATT\&CK as three distinct categories, focusing on how the attacker delivers the phishing attempt. This categorization supports our analysis of intent in phishing emails, particularly in the context of LLM-based detection. By emphasizing the delivery vector rather than attribution or payload analysis, our use of this taxonomy aligns with the goal of examining how LLMs can interpret the purpose behind an email. To generalize the classification and reflect the broader applicability to various phishing scenarios—including those involving LLM-generated content—we omit the term 'spear' from the category names, while preserving the core distinctions among the attack vectors. 

\textbf{Phishing via Link} refers to phishing emails designed to lure users into clicking on a link or visiting a website. Methods may include the use of shortened URLs, links that closely resemble legitimate domains but contain slight variations (e.g., a single altered character), or obfuscated, non-clickable links. For instance, a URL might be disguised using textual substitutions such as '(dot)com' in place of '.com' to deceive recipients into manually entering the address in their browser. Overall, this category encompasses all phishing attempts that seek to redirect users to malicious websites, whether through direct clicking or more indirect methods."

\textbf{Phishing via Attachment} refers to the method of delivering malicious code through a file that is attached to an email. This method relies on the victim downloading and interacting with the attachment to initiate a cyber infection. This category applies when a malicious file is attached, and the attacker aims for the victim to open it.  
\textit{It is important to note that the experiments conducted in this study focused solely on the text fields within the body and subject of the email. Consequently, the attachment was not included as part of the system's input. As a result, the system’s outcomes are based exclusively on the text fields without any access to the actual attachment.}

\textbf{Phishing via Service}  refers to a broader category of phishing attacks that utilize vectors outside the traditional email inbox, meaning the threat does not originate from a link or attachment within the email itself. Instead, attackers typically attempt to redirect the victim to engage through less secure and less monitored channels, such as personal phone numbers, SMS, or even physical mail. These emails usually contain just enough information to prompt the recipient to take further action, such as initiating a money transfer, installing software, or continuing the interaction through third-party services. This category highlights phishing techniques that exploit external communication channels to bypass conventional email-based defenses. 
\begin{table}[h]
    \caption{Transposing MITRE ATT\&CK Techniques to Phishing Categories}
    \label{tab:mitre_phishing_mapping}
    \centering
    \begin{tabular}{|c|c|}
        \hline
        \textbf{Technique} & \textbf{Phishing Category} \\  
        \hline
        T1566.001 Spearphishing Attachment &  Phishing via Attachment \\  
        \hline
        T1566.002 Spearphishing Link &  Phishing via Link \\  
        \hline
        T1566.003 Spearphishing via Service&  Phishing via Service \\  
        \hline
    \end{tabular}
\vspace{2pt}
\end{table}

\section{Experimental Setup{\tiny}}

\subsection{Data sources and curation}
The primary dataset used in the experiments consisted of emails manually selected from three publicly available large email datasets: LING, Nazario, and Enron.
The phishing emails were chosen from the LING and Nazario datasets, and legitimate emails were sourced from the Enron dataset. These datasets were selected due to their popularity and their compliance with privacy, especially with benign emails. The labeled datasets were downloaded from Kaggle \cite{PhishingEmailDataset2024}.

During initial experiments, it was observed that when using Enron emails that mention specific references to the company or its products, the LLMs sometimes recognized them as originating from the Enron dataset. Although interesting, this could shift the focus away from email intent and result in skewed results. To maintain concentration on detecting intent, such emails were filtered out.

After initial experiments, a validation set of 100 manually labeled emails was created to ensure an unbiased evaluation of classification and categorization during the final testing phase. This validation set adhered to the same labeling schema as the first dataset but remained unused until the end of the project to minimize any bias in training.

This research incorporated datasets with varying origins, sizes, and levels of complexity to provide a robust assessment of LLM capabilities in detecting phishing emails in the real world. 

\subsubsection{Data Preprocessing}
In order to standardize the data for analysis, we processed the data from the datasets in the following way:
\begin{enumerate}
    \item Extraction of email components: We extracted the text fields from all datasets, specifically the "Subject" field as the header and the "Body" field as the primary text content of each email.
    \item Binary label identification: From each dataset, the emails were labeled with a binary label, where a value of 1 indicated a phishing email, and 0 represented a legitimate message.
    \item Manual labeling and categorization: For the two custom datasets, all emails with the phishing label were manually categorized according to the corresponding intent categories from the taxonomy.
   \item Filtering out dataset bias: During experiments, some emails, like those from the Enron dataset, had clear indicators that caused the LLMs to recognize the text. For these cases and other examples, such as data formatting errors in the datasets, the emails were removed and replaced.
\end{enumerate}
\subsection{Prompting Approaches}
\subsubsection{Zero-shot Prompting}
In the zero-shot experiments, the prompts are constructed without providing specific examples of phishing or legitimate emails. Instead, they rely on descriptive guidance highlighting key features to identify. The classification prompt remains relatively simple, while the categorization prompt incorporates more detailed criteria. This zero-shot approach leverages the model's pre-trained knowledge by asking it to assess whether an email is malicious based solely on its internal understanding, without requiring explicit examples.

As a first step, the model is prompted with a binary (yes-or-no) question to determine whether the email is malicious. If the response is affirmative, the second step involves classifying the email into a \textit{intent} category, which reflects what the attacker aims to prompt the recipient to do. Since no examples are provided, the model must rely entirely on its pre-trained knowledge to infer the characteristics of a malicious email and its underlying intent.

\subsubsection{Few-shot Learning Prompt}
To enhance accuracy, particularly in categorization, we implemented a few-shot learning prompting approach. The main distinction in the few-shot prompt is that each category includes two complete examples of phishing emails, encompassing both the header and body. This approach provides the models with real examples as references, aiming to improve their ability to recognize various patterns within the categories.

\subsection{The experiments}
This project conducted three experiments, all of which employed two different approaches: zero-shot and few-shot. All emails were sent in single prompts, with no overlap between each model; each model received only one prompt at a time to avoid data contamination.

All experiments are first conducted using a \textbf{zero-shot approach} by sending only the prompt with the email without any examples. The LLMs execute their task by using only the prompt's instructions and their own capabilities.

In the \textbf{few-shot approach}, all experiments are conducted using the same base prompt, that base prompt is augmented in each experiment. In the few-shot version of the experiments the prompt is further augmented with two labeled examples for each category of phishing emails relevant to the specific experiment. These examples serve as contextual cues to help the LLM learn from limited data—a technique commonly referred to as few-shot prompting. This setup allows us to investigate the model’s ability to generalize and classify phishing intent based on minimal, targeted guidance. 

\textbf{Experiment 1} investigates the LLM's innate knowledge by using basic prompt engineering and asking if the email is malicious. A commonly used prompt engineering technique, often referred to as "role prompting," is employed to convey the purpose that the LLM will be utilized as an email classifier in the initial step. The experiment introduces the prompt shown below, which will serve as a baseline for all the other experiments. The experiment was run twice, using the same prompt, with the difference that the few-shot examples of emails and the desired output were appended at the end of the prompt. The few-shot prompt in experiment 1 does not include the intent categories. 
\lstset{
  basicstyle=\ttfamily\small,
  frame=single,
  breaklines=true
}

\begin{lstlisting}
You are an email classifier analyzing potential phishing emails. Your task is as follows:

1. Determine if this email is malicious (Yes/No) 
2. Give a short justification for your decision, explain the result.

The response should follow this format

Phishing: YES/NO
Justification: 
\end{lstlisting}
\textbf{Experiment 2} enhances the prompt by introducing intent categories in Step 1. This addition provides the LLM with more contextual information but does not constitute an additional step in the overall process. These intent categories are also included in the few-shot learning examples to guide the model more effectively.

\begin{lstlisting}
 1.Determine if this email is malicious (Yes/No).  
Here are a few categories of phishing emails and some basic rules on how to find them :
 - Phishing via Attachment
    If the primary goal of the phishing email is to get the user to download something
   - Phishing via Link
    If the primary goal of the phishing email is to get the user to "click here" or click any URL or link
   - Phishing via Service
    Where the goal is not To CLICK or download in the inbox, but to get the user to use some other service, like calling a number or some other way they could phish outside of the email inbox
   - Other
   If the email is clearly a phishing attempt but does not fall into any of the defined categories
\end{lstlisting}

\textbf{Experiment 3} incorporates all three steps, building on the initial assessment of the LLM’s capability by introducing a second step: a categorization task. This expanded approach aims to evaluate the LLM’s ability to perform a more comprehensive analysis, extending beyond simple binary classification, by focusing on its understanding of various phishing tactics and its reasoning capabilities.

\begin{lstlisting}
1. Determine if this email is malicious (Yes/No).
2. ONLY If the email is malicious, classify it into one of the following categories:
  - Phishing via Attachment
    If the primary goal of the phishing email is to get the user to download something
   - Phishing via Link
    If the primary goal of the phishing email is to get the user to "click here" or click any URL or link
   - Phishing via Service
    Where the goal is not To CLICK or download in the inbox, but to get the user to use some other service, like calling a number or some other way they could phish outside of the email inbox
   - Other
   If the email is clearly a phishing attempt but does not fall into any of the defined categories
3. Give a short justification for your decision, and explain the result.
\end{lstlisting}

\subsection{Model selection}
The experiments utilized four models: GPT-4o-mini, Claude 3.5 Haiku, Phi-4 (14B), and Qwen (7B). The objective was not to determine the most capable model, but rather to explore the effectiveness of modern large language models in phishing detection and categorization. Qwen (7B), the smallest and oldest model (over a year old), was included to evaluate how a smaller, less recent model performs in comparison to newer, larger, and more cost-effective enterprise models. Claude 3.5 Haiku and GPT-4o-mini were accessed via commercial APIs, while Qwen (7B) and Phi-4 (14B) were run locally on a high-end consumer desktop.

\section{Results}
The experiments progressed through three stages: (1) basic malicious email identification (Exp1); incorporation of phishing technique categorization (Exp2); and (3) combining both tasks with an added justification requirement (Exp3). For each stage, we used both zero-shot and few-shot learning approaches, suffixed by 
`-Zero' and `-Few' respectively in Table \ref{tab:ieee_table_compact}, which summarizes the results.

\begin{table}[H]
\caption{Accuracy across experiments. Category accuracy is shown as \textit{Detection / Category} where applicable.}
\label{tab:ieee_table_compact}
\centering
\scriptsize
\setlength{\tabcolsep}{7pt}  
\renewcommand{\arraystretch}{1.0}  
\begin{tabular}{lccc}
\toprule
\textbf{Model} & {Exp1-Zero} & {Exp1-Few} & {Exp2-Zero} \\
\midrule
gpt-4o-mini      & 97.00\% & 97.00\% & 93.00\% \\
claude-3.5-haiku & 96.00\% & 92.00\% & 95.00\% \\
phi-4(14b)           & 90.00\% & 92.00\% & 91.00\% \\
qwen(7b)             & 44.00\% &  2.00\% & 45.00\% \\
\midrule
\textbf{Model} & {Exp2-Few} & {Exp3-Zero} & {Exp3-Few} \\
\midrule
gpt-4o-mini      & 92.00\% & 94.00\% / 86.05\% & 92.00\% / 95.35\% \\
claude-3.5-haiku & 92.00\% & 89.00\% / 88.37\% & 94.00\% / 79.07\% \\
phi-4(14b)            & 88.00\% & 93.00\% / 86.05\% & 89.00\% / 76.74\% \\
qwen(7b)             &  0.00\% & 25.00\% / 9.30\%   &  0.00\% / 0.00\%   \\
\bottomrule
\end{tabular}
\vspace{2pt}
\end{table}
Across all experiments, GPT-4o-mini, Claude-3.5-haiku, and Phi-4 (14b) consistently demonstrated high accuracy, highlighting their ability to understand and classify malicious emails even with limited example data. The Qwen(7b) performed considerably worse than the other models. In some tasks, it failed to produce output in the correct format, resulting in zero percentage accuracy. The inclusion of categorization focuses on what the attacker intends for the targeted user to perform, which could give security professionals a head start in the triage process of a real attack. The requirement for justification provides some insight into the model's reasoning process and transparency. The complete suite of six experiments, when executed in a single batch, requires approximately 70 minutes to complete. The overall execution time is primarily constrained by the locally hosted models. In contrast, experiments conducted exclusively via API access typically take between 1 to 3 minutes per experiment, incurring a cost of approximately \$0.01 to \$0.03 USD for the GPT-4o-mini and Claude Haiku models.

All experiments also required the models to generate justifications as part of the output. Consistent with the results from Steps 1–3, Qwen exhibited inadequate performance on this task. Additionally, Phi-4 and Claude encountered formatting issues that led to empty justifications in up to one-third of the emails. These shortcomings indicate clear opportunities for improvement in the justification generation process. The justifications provided in the correct format were of high quality and offered good logic for determining whether an email appeared legitimate or suspicious. Example justifications for both a legitimate and a phishing email via a link are included below:\\

Legitimate email:
\begin{lstlisting}
The email appears to be a legitimate inquiry about linguistic analyses and does not contain any malicious intent, links, or attachments that would indicate phishing. It is a straightforward request for information from a researcher.
\end{lstlisting}

Phishing via Link:
\begin{lstlisting}
The email is attempting to get the recipient to click on a link to verify their account, which is a common tactic used in phishing attempts. The urgency created by the threat of account suspension within 24 hours further indicates malicious intent.
\end{lstlisting}

\section{Conclusion and Future work}

In this paper, we have evaluated the potential use of LLMs to detect and categorize phishing emails based on the attacker's \textit{intent}. The experimental results demonstrated that modern LLM models are capable of inferring attack vectors aligned with the categories in the proposed taxonomy. Furthermore, the outputs generated by these models offer valuable insights for security professionals. Returning to the research questions outlined in Section \ref{sec:intro}, our findings can be summarized as follows:

\textbf{(RQ1)} LLMs, particularly the larger modern models, demonstrated a strong ability to infer phishing intent, achieving above 95\% accuracy in phishing detection. This extends beyond keyword spotting, as models accurately identified how emails attempted to trick the user by focusing on the attacker's intent.

\textbf{(RQ2)} LLMs exhibited substantial inherent knowledge, achieving high accuracy in zero-shot experiments. However, incorporating two examples per category via few-shot learning had mixed results: for some models, the examples improved the accuracy of categories, while in other cases, we observed a reduction in accuracy. Context length and the models' size might be important if few-shot learning is to be introduced or further explored.

\textbf{(RQ3)} Justification alongside detection and categorization results revealed insight into the model's thought process. While a comprehensive analysis of reasoning quality was beyond the scope of this work, a preliminary examination of the generated justifications showed a correlation between
identified phishing cues and the model's stated rationale. This suggests the LLMs weren't simply relying on surface-level features but were, to some degree, able to connect the intent of the phishing email to the categories.

\textbf{(RQ4)} LLMs successfully categorized phishing emails into distinct categories (i.e., link, attachment, service). Category accuracy ranged from 76\% to 95\% on the three best-performing models (see Table II), indicating that based only on the text of the email, the LLMs can sort the emails into spearphishing techniques with high accuracy. This demonstrates a significant ability to differentiate between phishing categories, moving beyond simple binary classification.

\textbf{(RQ5)} By categorizing phishing emails, LLMs demonstrate the ability to leverage domain-specific knowledge to identify the attacker’s intent, classify and categorize the threat, and generate explanatory justifications for why the email is deemed malicious. The extracted information—including identified indicators and inferred intent—can complement traditional security filters, particularly in cases where phishing emails bypass existing detection systems. This added layer of analysis has the potential to reduce false positives and assist security professionals in the triage and investigation of email-based security incidents.

Based on our work, we have identified several areas for future research. Firstly, as discussed, we suggest addressing reasoning and justification capabilities (RQ3) in a controlled experiment with security analysts. Other areas for further work include: evaluating the models on a larger and more diverse real-world dataset of phishing emails; investigating the integration of an LLM-based approach with existing email security systems; fine-tuning of existing LLMs; conducting a cost-benefit analysis of the use of LLMs; exploring different prompting strategies; and implementing a human-in-the-loop validation system. These future research directions could contribute to a deeper understanding of the capabilities and limitations of LLMs for phishing detection and categorization.

\section*{Acknowledgment}
The authors wish to express their sincere gratitude to Mr. Mateusz Zych for his thoughtful feedback and constructive suggestions on the manuscript.



\begin{thebibliography}{00}

\bibitem{rekouche2011early}
K. Rekouche, ``Early phishing,'' *arXiv preprint arXiv:1106.4692*, 2011.

\bibitem{alkhalil2021phishing}
Z. Alkhalil, C. Hewage, L. Nawaf, and I. Khan, "Phishing attacks: A recent comprehensive study and a new anatomy," *Frontiers in Computer Science*, vol. 3, p. 563060, 2021.

\bibitem{cofense2024emailsecurity}
Cofense, "2024 Cofense Annual State of Email Security Report," Cofense, Feb. 2024. [Online]. Available: \url{https://cofense.com/getmedia/db5a5ad7-b39a-45f5-bab7-eb165b9a0685/2024-cofense-annual-state-of-email-security-report.pdf}. [Accessed: Apr. 25, 2025].

\bibitem{microsoft2024digitaldefense}
Microsoft, "Microsoft Digital Defense Report 2024,'' Microsoft, Oct. 2024. [Online]. Available: \url{https://cdn-dynmedia-1.microsoft.com/is/content/microsoftcorp/microsoft/final/en-us/microsoft-brand/documents/Microsoft%20Digital%20Defense%20Report%202024%20%281%29.pdf}. [Accessed: Apr. 25, 2025].

\bibitem{MITRE_ATTCK}
MITRE Corporation, "MITRE ATT\&CK: Phishing (T1566)," [Online]. Available: \url{https://attack.mitre.org/techniques/T1566/}. [Accessed: Feb. 28, 2025].

\bibitem{brickley2021comparative}
J. C. Brickley, K. Thakur, and A. S. Kamruzzaman, "A comparative analysis between technical and non-technical phishing defenses," *Int. J. Cyber-Security and Digital Forensics*, vol. 10, no. 1, pp. 28--41, 2021.

\bibitem{durumeric2015neither}
Z. Durumeric *et al.*, "Neither snow nor rain nor MITM... an empirical analysis of email delivery security," in *Proc. 2015 Internet Measurement Conf.*, pp. 27--39, 2015.

\bibitem{hazell2023large}
J. Hazell, "Large language models can be used to effectively scale spear phishing campaigns," *arXiv preprint arXiv:2305.06972*, 2023.

\bibitem{Heiding2024devising}
F. Heiding, B. Schneier, A. Vishwanath, J. Bernstein, and P. S. Park, "Devising and detecting phishing emails using large language models," *IEEE Access*, 2024.

\bibitem{derouet2016fighting}
E. Derouet, "Fighting phishing and securing data with email authentication," *Computer Fraud \& Security*, vol. 2016, no. 10, pp. 5--8, 2016.

\bibitem{koide2024chatspamdetector}
T. Koide, N. Fukushi, H. Nakano, and D. Chiba, "Chatspamdetector: Leveraging large language models for effective phishing email detection," *arXiv preprint arXiv:2402.18093*, 2024.

\bibitem{yao2024survey}
Y. Yao, J. Duan, K. Xu, Y. Cai, Z. Sun, and Y. Zhang, "A survey on large language model (LLM) security and privacy: The good, the bad, and the ugly," *High-Confidence Computing*, p. 100211, 2024.

\bibitem{jampen2020don}
D. Jampen, G. Gür, T. Sutter, and B. Tellenbach, "Don't click: Towards an effective anti-phishing training. A comparative literature review," *Human-centric Computing and Information Sciences*, vol. 10, no. 1, p. 33, 2020.

\bibitem{kojima2022large}
T. Kojima, S. S. Gu, M. Reid, Y. Matsuo, and Y. Iwasawa, "Large language models are zero-shot reasoners," *Advances in Neural Information Processing Systems*, vol. 35, pp. 22199--22213, 2022.

\bibitem{PhishingEmailDataset2024}
N. A. Alam, "Phishing Email Dataset," [Online]. Available: \url{https://www.kaggle.com/datasets/naserabdullahalam/phishing-email-dataset}. [Accessed: Feb. 20, 2025].

\bibitem{mitreT1566_002}
MITRE, "Phishing: Spearphishing Link," [Online]. Available: \url{https://attack.mitre.org/techniques/T1566/002/}. [Accessed: Apr. 3, 2025].

\end{thebibliography}
\end{document}